\documentclass{article}
\usepackage{spconf,amsmath,graphicx}
\usepackage{cite}
\usepackage{makecell}
\usepackage[squaren]{SIunits}
\usepackage{multirow}
\usepackage[table]{xcolor}%
\usepackage{hhline}
\usepackage[acronym]{glossaries}
\usepackage{todonotes}
\usepackage{support-caption}
\usepackage{subcaption}
\usepackage[inkscape=false]{svg}
\usepackage{url}
\usepackage{soul}
\usepackage{algorithm}
\usepackage{algpseudocode}
\usepackage{enumitem}
\usepackage{lipsum}
\usepackage{booktabs}
\usepackage{pdfpages}
\usepackage{hyperref}
\usepackage{amsfonts}

\makeatletter
\def\bstctlcite{\@ifnextchar[{\@bstctlcite}{\@bstctlcite[@auxout]}}
\def\@bstctlcite[#1]#2{\@bsphack
  \@for\@citeb:=#2\do{%
    \edef\@citeb{\expandafter\@firstofone\@citeb}%
    \if@filesw\immediate\write\csname #1\endcsname{\string\citation{\@citeb}}\fi}%
  \@esphack}
\makeatother

\makeatletter
\newcommand\notsotiny{\@setfontsize\notsotiny\@vipt\@viipt}
\makeatother

\newacronym{2d}{2D}{two-dimensional}

\newacronym{asr}{ASR}{automatic speech recognition}
\newacronym{nmt}{NMT}{neural machine translation}
\newacronym{tts}{TTS}{text-to-speech}

\newacronym{brir}{BRIR}{binaural room impulse response}

\newacronym{dnn}{DNN}{deep neural network}
\newacronym{lstm}{LSTM}{long short-term memory}

\newacronym{ic}{IC}{interaural coherence}
\newacronym{ild}{ILD}{interaural level difference}
\newacronym{irm}{IRM}{ideal ratio mask}
\newacronym{itd}{ITD}{interaural time difference}

\newacronym{logfbe}{log-FBE}{log-filterbank energy}

\newacronym{mse}{MSE}{mean squared error}

\newacronym{pesq}{PESQ}{perceptual evaluation of speech quality}

\newacronym{relu}{ReLU}{rectified linear unit}
\newacronym{rms}{RMS}{root mean square}

\newacronym{snr}{SNR}{signal-to-noise ratio}
\newacronym{stoi}{STOI}{short-term objective intelligibility}
\newacronym{estoi}{ESTOI}{extended short-term objective intelligibility}
\newacronym{sisnr}{SI-SNR}{scale-invariant signal-to-noise ratio}

\newacronym{tf}{T-F}{time-frequency}

\newacronym{zpr}{ZPR}{zero-padding rate}

\def\dpesq{\Delta\text{\gls{pesq}}}
\def\destoi{\Delta\text{\gls{estoi}}}
\def\dsnr{\Delta\text{\gls{snr}}}

\title{On batching variable size inputs for training end-to-end speech enhancement systems}

\name{Philippe Gonzalez$^{\star}$, Tommy Sonne Alstrøm$^{\dagger}$, Tobias May$^{\star}$}

\address{$^{\star}$Department of Health Technology, Technical University of Denmark\\$^{\dagger}$Department of Applied Mathematics and Computer Science, Technical University of Denmark}

\begin{document}
\bstctlcite{IEEEexample:BSTcontrol}

\ninept

\maketitle

\begin{abstract}
The performance of neural network-based speech enhancement systems is primarily influenced by the model architecture, whereas training times and computational resource utilization are primarily affected by training parameters such as the batch size. Since noisy and reverberant speech mixtures can have different duration, a batching strategy is required to handle variable size inputs during training, in particular for state-of-the-art end-to-end systems. Such strategies usually strive for a compromise between zero-padding and data randomization, and can be combined with a dynamic batch size for a more consistent amount of data in each batch. However, the effect of these strategies on resource utilization and more importantly network performance is not well documented. This paper systematically investigates the effect of different batching strategies and batch sizes on the training statistics and speech enhancement performance of a Conv-TasNet, evaluated in both matched and mismatched conditions. We find that using a small batch size during training improves performance in both conditions for all batching strategies. Moreover, using sorted or bucket batching with a dynamic batch size allows for reduced training time and GPU memory usage while achieving similar performance compared to random batching with a fixed batch size.
\end{abstract}


\section{Introduction}
\label{sec:intro}
The performance of deep learning-based speech enhancement systems has drastically improved in recent years, in particular if they are trained with very large datasets \cite{chen2016large}. However, this substantially increases training times.
While most of the recent research in speech enhancement has focused on designing new architectures \cite{wang2018supervised,lv2021dccrnplus,pandey2022tparn}, little research has been dedicated to designing efficient data loading pipelines that optimize training times. The usual method to reduce training time consists in tweaking training parameters such as the learning rate or the batch size. However, for applications with variable size inputs such as \gls{asr}, \gls{tts}, \gls{nmt}, source separation or speech enhancement, forming batches from training examples is not as simple as stacking along a new axis, as this would result in non-contiguous tensors.
State-of-the-art end-to-end systems explicitly exploit temporal context via the use of e.g.\ convolutional layers \cite{lv2021dccrnplus} or recurrent architectures \cite{pandey2022tparn}, preventing from defining a batch as multiple time samples of feature frames.
A common solution to obtain contiguous batches of variable size inputs is to zero-pad the sequences within a batch to match their lengths. However, this increases the amount of data passed through the network, leading to computational waste and increased training times.
Moreover, increasing the batch also increases the total amount of zero-padding required, and more importantly seems to reduce the generalization performance of deep neural networks \cite{keskar2016large}.

One solution to minimize the amount of zero-padding is to sort the training sequences by length before forming the batches. However, a common assumption in stochastic gradient descent is the need of i.i.d. sampling of the training data to ensure unbiased gradient estimators \cite{bengio2012practical}. When sorting the sequences, the same sequences are batched together in every training epoch, even though the batches can be shuffled subsequently. Thus, the reduction in zero-padding comes at the expense of data randomization. Batching strategies with a tunable parameter defining this compromise include bucket batching \cite{khomenko2016accelerating,variani2017end}, alternated-sorting batching \cite{doetsch2017comprehensive} and semi-sorted batching \cite{ge2021speed}. Doetsch et al. \cite{doetsch2017comprehensive} conducted a comparative study of random batching, sorted batching, bucket batching and semi-sorted batching for the training of a \gls{lstm}-based \gls{asr} system. They showed that different batching strategies can substantially influence training times and speech recognition performance. Ge et al. \cite{ge2021speed} introduced semi-sorted batching for the training of a Tacotron \cite{shen2018natural}-based \gls{tts} system, and showed that bucket batching, alternated-sorting batching and semi-sorted batching performed similarly and provided significant training time improvements compared to random batching. To the best of our knowledge, the influence of different batching strategies on speech enhancement performance has not been systematically investigated. In particular, the effect of different batching strategies on generalization performance is unknown. Moreover, recent studies in distributed optimization have questioned the need for i.i.d. sampling of the training data \cite{meng2017convergence, nguyen2022why}, hinting towards the viability of batching strategy alternatives to the commonly used random batching.

Another consideration when constructing batches of variable size inputs is whether to define a dynamic batch size. Indeed, stacking a fixed amount of variable length sequences results in batches of variable total size, potentially leading to numerically unstable gradients and inefficient GPU memory allocation. A solution is to use a dynamic batch size in terms of e.g.\ a total number of tokens in the case of \gls{nmt}, or a total duration in seconds in the case of speech processing. Morishita et al. \cite{morishita2017empirical} compared different fixed and dynamic batch sizes in the context of \gls{nmt} using a \gls{lstm}-based model. They concluded that using a fixed or dynamic batch size has little effect on performance, and that the batch size can substantially affect the final accuracy of the model. However, they did not look at the effect on generalization performance.

The present study compares three batching strategies, namely random, sorted and bucket batching, in terms of training statistics (training time, allocated GPU memory and padding amount) and speech enhancement performance of a Conv-TasNet \cite{luo2019conv} in both matched and mismatched conditions. For each batching strategy, the effect of the batch size as well as its nature (fixed or dynamic) is investigated. In Sec.~\ref{sec:batching}, we provide an overview of the evaluated batching strategies. In Sec.~\ref{sec:system}, we describe the system used to perform the speech enhancement task. In Sec.~\ref{sec:setup}, we describe the experimental setup used to train and test the model with the different batching strategies. Finally in Sec.~\ref{sec:results}, we report the results in terms of training statistics and speech enhancement performance in both matched and mismatched conditions.

\section{Batching strategies}
\label{sec:batching}
\begin{figure}
  \footnotesize
  \centering
  \includesvg{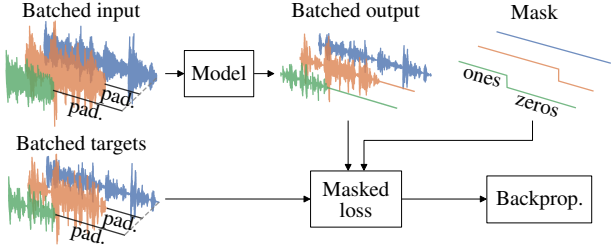}
  \vspace{-2pt}
  \caption{Batching process. Input mixtures are padded to match their length and passed as a contiguous tensor through the network. A masked loss ensures the padded regions do not contribute to updating the weights.}
  \label{fig:draw}
\end{figure}
\begin{figure}[t!]
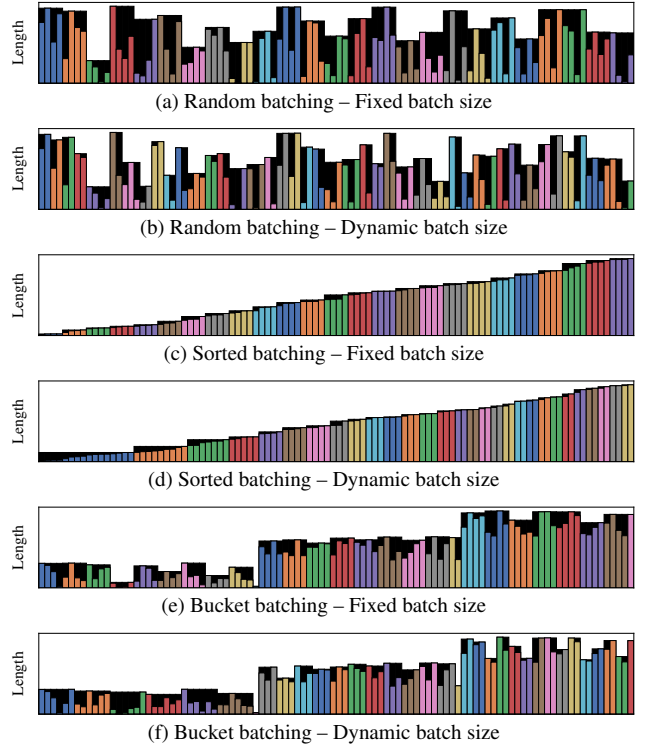

  \captionsetup[subfigure]{aboveskip=2pt, belowskip=-3pt}
  \notsotiny
  \centering
  \begin{subfigure}{\linewidth}
    \centering
    \includesvg[scale=0.97]{pics/batching_random_fixed.svg}
    \caption{Random batching -- Fixed batch size}
    \label{fig:batching_random_fixed}
  \end{subfigure} \\[7pt]
  \begin{subfigure}{\linewidth}
    \centering
    \includesvg[scale=0.97]{pics/batching_random_dynamic.svg}
    \caption{Random batching -- Dynamic batch size}
    \label{fig:batching_random_dynamic}
  \end{subfigure} \\[7pt]
  \begin{subfigure}{\linewidth}
    \centering
    \includesvg[scale=0.97]{pics/batching_sorted_fixed.svg}
    \caption{Sorted batching -- Fixed batch size}
    \label{fig:batching_sorted_fixed}
  \end{subfigure} \\[7pt]
  \begin{subfigure}{\linewidth}
    \centering
    \includesvg[scale=0.97]{pics/batching_sorted_dynamic.svg}
    \caption{Sorted batching -- Dynamic batch size}
    \label{fig:batching_sorted_dynamic}
  \end{subfigure} \\[7pt]
  \begin{subfigure}{\linewidth}
    \centering
    \includesvg[scale=0.97]{pics/batching_bucket_fixed.svg}
    \caption{Bucket batching -- Fixed batch size}
    \label{fig:batching_bucket_fixed}
  \end{subfigure} \\[7pt]
  \begin{subfigure}{\linewidth}
    \centering
    \includesvg[scale=0.97]{pics/batching_bucket_dynamic.svg}
    \caption{Bucket batching -- Dynamic batch size}
    \label{fig:batching_bucket_dynamic}
  \end{subfigure}
  \caption{The different batching strategies. Adjacent observations of the same color are batched together. Zero-padding is represented in black. The number of buckets was set to 3 in (e) and (f).}
  \label{fig:batching}
\end{figure}

The batching task is illustrated in Fig.~\ref{fig:draw} and can be described as follows. $N$ noisy and reverberant mixtures ${x_1,\ldots,x_N}$ with respective lengths ${T_1,\ldots,T_N}$ are padded such that they can be stacked to form a contiguous batch in $\mathbb{R}^{N\times T_{\max}}$, where $T_{\max}=\max\{T_1,\ldots,T_N\}$. The batch is then forward passed through the network and the output has the same dimensions as the input batch. Since the output batch also contains the model output corresponding to the padded zeros, a masked loss must be used to prevent the zero-padded regions from contributing to the loss.
To reduce the amount of padding and thus computational waste, strategies can be used to create batches from all the sequences in the training dataset. In the present study, we consider the three strategies illustrated in Fig.~\ref{fig:batching}:
\begin{itemize}[leftmargin=10pt]
\item \textit{Random batching}: at the start of each epoch, the sequences are randomized before creating and shuffling the batches. This corresponds to the usual practice in deep learning. While this ensures randomization of sequences during training, it requires the largest amount of zero-padding compared to sorted and bucket batching, as indicated by the black regions in Fig.~\ref{fig:batching}.
\item \textit{Sorted batching}: before training, the sequences are sorted by their length before creating the batches. This allows to batch sequences of similar length together, which in turn minimizes the amount of padding, as seen in Fig.~\ref{fig:batching}. The batches are then shuffled at the start of each epoch. This reduces computational waste at the expense of randomization, since the same observations are batched together in each epoch.
\item \textit{Bucket batching}: available in TensorFlow \cite{tensorflow2015,variani2017end} and described in \cite{khomenko2016accelerating}, this strategy can be seen as a compromise between random and sorted batching. Before training, sequences are assigned to different buckets according to their length, such that sequences of similar length are in the same bucket. Then, at the start of each epoch, batches are created using random sequences from the same bucket. The number of buckets defines a compromise between randomization and batching; using a single bucket is the same as random batching, while using as many buckets as the number of batches is the same as sorted batching. The bucket limits can be uniformly spaced between the minimum and maximum sequence length, or defined as quantiles of the sequence length distribution such that each bucket contains the same number of sequences.
\end{itemize}

For each batching strategy, a fixed or a dynamic batch size can be used. A fixed batch size is defined as a fixed number of sequences as typically done in deep learning, while a dynamic batch size is defined as a total length of sequences in the batch after padding. This is particularly useful when the sequence length distribution is broad, in which case using a fixed batch size would result in a highly variable amount of data in each batch. A dynamic batch size allows batches of short sequences to contain more sequences, while batches of long sequences to contain fewer sequences.

\section{Speech enhancement system}
\label{sec:system}

The speech enhancement system takes as input a noisy and reverberant binaural mixture sampled at 16\,kHz. Binaural mixtures are created by convolving a clean speech utterance and noise recordings with \glspl{brir} from the same room.
The binaural mixture is averaged across left and right channels before being fed to a Conv-TasNet \cite{luo2019conv}, which is an end-to-end fully convolutional network designed for single-channel multi-speaker speech separation. It can be used for speech enhancement as described in \cite{koyama2020exploring} by optimizing the \gls{snr} loss instead of the \gls{sisnr} \cite{leroux2019sdr} loss, and by setting the number of separated sources to $K=1$. The target source is defined as the direct-sound part of the speech signal including early reflections up to a boundary of 50\,ms, which was shown to be beneficial for speech intelligibility \cite{roman2013speech}. This is done by splitting the \gls{brir} used to generate the mixture into a direct-sound part and a reverberant part using a windowing procedure as described in \cite{zahorik2002direct}. Discarding late speech reflections from the target signal also allows to train the Conv-TasNet to perform dereverberation.
We use one of the smaller configurations proposed in \cite{luo2019conv} while keeping a filter length of 2\,ms at 16\,kHz, namely $N=128$, $L=32$, $B=128$, $H=256$, $S_c=128$, $P=3$, $X=7$ and $R=2$. We use cumulative layer normalization (cLN) to make the system causal as described in \cite{luo2019conv}. The model is trained for 150 epochs using the Adam optimizer \cite{kingma2014adam} with a learning rate of $1e^{-3}$. Gradients are clipped with a maximum $L_2$-norm of 5. The network has $1.45\,\text{M}$ parameters.

\section{Experimental setup}
\label{sec:setup}

For each batching strategy, the speech enhancement system is trained using different fixed or dynamic batch sizes. The considered fixed batch size are 1, 2, 4 and 8 sequences, while the considered dynamic batch sizes are 2\,s, 4\,s, 8\,s, 16\,s, 32\,s, 64\,s and 128\,s. For bucket batching, we use 10 buckets and uniformly spaced bucket limits between the minimum and maximum mixture length. For each configuration, the model is trained five times using different random seeds for weight initialization, batch generation and shuffling. All trainings are performed on machines equipped with Intel Xeon Gold 6126 CPUs and Tesla V100 GPUs.

\begin{figure}
  \footnotesize
  \centering
  \includesvg[width=.99\linewidth]{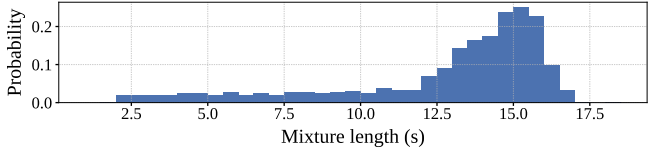}
  \vspace{-6pt}  
  \caption{Training mixture length distribution.}
  \label{fig:dist}
\end{figure}

The model is trained on 10~hours of noisy and reverberant mixtures. Clean speech utterances were selected from LibriSpeech (100-hour version) \cite{panayotov2015librispeech}, noise recordings from the TAU urban acoustic scenes 2019 development dataset \cite{heittola2019tau}, and \glspl{brir} from the university of Surrey \cite{hummersone2010surrey}. The \gls{snr} is uniformly distributed between -5\,dB and 10\,dB. The target speech and up to 3 noise sources are randomly placed in the horizontal plane between -90$\degree$ and 90$\degree$ in front of the receiver. Figure~\ref{fig:dist} shows the training mixture length distribution. When the batch size is dynamic and smaller than the maximum mixture length, the mixtures are split into segments shorter than the batch size such that a batch fits at least one segment.

The system is evaluated in matched and mismatched conditions. The matched condition consists of mixtures generated from different speech utterances, noise recordings and \glspl{brir}, but drawn from the same databases as during training. The mismatched conditions consist of mixtures generated from different speech corpora, noise databases and \gls{brir} databases. The speech corpora are TIMIT \cite{garofolo1993timit}, WSJ SI-84 \cite{paul1992design}, Clarity \cite{cox2022clarity} and VCTK \cite{yamagishi2019vctk}. The noise databases are NOISEX\cite{varga1993noisex}, ICRA \cite{dreschler2001icra}, DEMAND \cite{thiemann2013demand} and ARTE \cite{buchholz2019arte}. Finally the \gls{brir} databases are ASH \cite{shanon2021ash}, BRAS \cite{brinkmann2021bras}, CATT \cite{catt_brirs} and AVIL \cite{mccormack2020higher}. For each combination of databases, 1~hour of mixtures are generated for evaluation. The same random \gls{snr} range and positions in the horizontal plane as in the training set are used.

The system is evaluated in terms of \gls{pesq} \cite{recommendation2001perceptual}, \gls{estoi} \cite{jensen2016algorithm} and \gls{snr} improvements. The reference signal for each metric is the direct-sound part of the speech signal including early reflections as described in Sec.~\ref{sec:system}. The improvements between the unprocessed input mixture and the enhanced output signal are denoted as $\dpesq$, $\destoi$ and $\dsnr$. For each configuration, we also calculate the \gls{zpr}, which is defined as the total amount of padded zeros relative to the original dataset length, 
\begin{equation}
  \text{ZPR} = \frac{\sum_i^{N} p_i}{\sum_i^N l_i},
\end{equation}
where $N$ is the number of sequences in the dataset, $l_i$ is the length of the $i$-th sequence before padding and $p_i$ is the amount of zeros padded to the $i$-th sequence.

\section{Results}
\label{sec:results}

\begin{figure}
  \footnotesize
  \centering
  \includesvg[width=\linewidth]{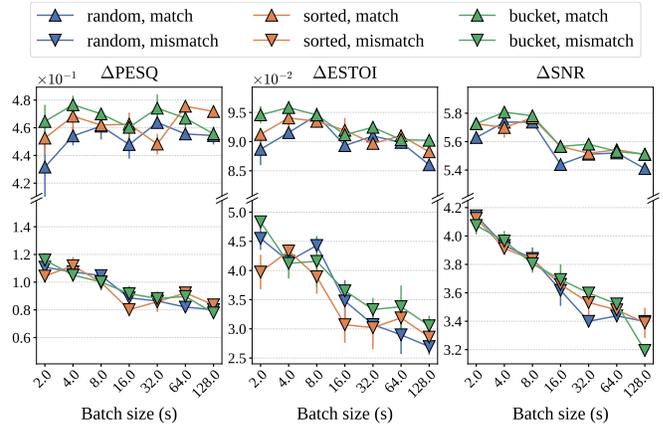}
  \vspace{-12pt}
  \caption{Speech enhancement performance in terms $\dpesq$, $\destoi$ and $\dsnr$ as a function of the dynamic batch size.}
  \label{fig:results}
\end{figure}

Table~\ref{tab:batching} shows the average $\dpesq$, $\destoi$ and $\dsnr$ scores obtained by the Conv-TasNet when trained with the different batching strategies and batch sizes. The standard error of the mean across random training seeds is displayed next to each average score. The table also displays different training statistics, namely the training time, the allocated GPU memory and the \gls{zpr}. As expected, the largest \gls{zpr} is obtained with random batching at large batch sizes, namely 24.5\% with a fixed batch size of 8 sequences and 24.1\% with a dynamic batch size of 128\,s. Conversely, sorted batching yields almost no padding even at large batch sizes. Bucket batching serves as a compromise between the two as we obtain a \gls{zpr} of 5.2\%. These differences at large batch sizes (when the data loading pipeline is not a bottleneck) are reflected in the training time; with a batch size of 128\,s, using random batching takes {7\,h\,18\,m}, while using sorted of bucket batching takes {5\,h\,48\,m} and {6\,h\,10\,m} respectively (a speed improvement of 26\% and 18\% respectively).

\begin{table*}[t!]
\footnotesize
\centering
\begin{tabular}{crrrrcccccc}
\hline \hline
&&&&&\multicolumn{3}{c}{Match}&\multicolumn{3}{c}{Mismatch}\\
\cmidrule(lr){6-8} \cmidrule(lr){9-11}
Strategy & \multicolumn{1}{c}{Batch size} & \multicolumn{1}{c}{Time} & \multicolumn{1}{c}{Memory} & \multicolumn{1}{c}{ZPR} & \makecell{$\Delta$PESQ\\[-2pt]{$\scriptstyle(\times 10^{-1})$}} & \makecell{$\Delta$ESTOI\\[-2pt]{$\scriptstyle(\times 10^{-2})$}} & $\Delta$SNR & \makecell{$\Delta$PESQ\\[-2pt]{$\scriptstyle(\times 10^{-1})$}} & \makecell{$\Delta$ESTOI\\[-2pt]{$\scriptstyle(\times 10^{-2})$}} & $\Delta$SNR \\
\hline \hline
\multirow{10}{*}{\rotatebox[origin=c]{90}{Random}}
& 1 seq. & 8\,h\,33\,m & 2.9\,GB & 0.0\% & \textbf{4.61} {\notsotiny $\pm$ \textbf{0.10}} & \textbf{9.02} {\notsotiny $\pm$ \textbf{0.12}} & \textbf{5.54} {\notsotiny $\pm$ 0.01} & \textbf{0.89} {\notsotiny $\pm$ 0.01} & \textbf{3.25} {\notsotiny $\pm$ 0.17} & \textbf{3.61} {\notsotiny $\pm$ 0.06} \\
& 2 seq. & 10\,h\,24\,m & 5.8\,GB & 14.2\% & 4.55 {\notsotiny $\pm$ 0.01} & 8.85 {\notsotiny $\pm$ 0.11} & 5.51 {\notsotiny $\pm$ 0.02} & 0.86 {\notsotiny $\pm$ 0.03} & 2.91 {\notsotiny $\pm$ 0.15} & 3.54 {\notsotiny $\pm$ \textbf{0.08}} \\
& 4 seq. & 8\,h\,05\,m & 11.6\,GB & 21.0\% & 4.56 {\notsotiny $\pm$ 0.08} & 8.67 {\notsotiny $\pm$ 0.06} & 5.46 {\notsotiny $\pm$ \textbf{0.04}} & 0.82 {\notsotiny $\pm$ 0.03} & 2.88 {\notsotiny $\pm$ 0.25} & 3.48 {\notsotiny $\pm$ 0.06} \\
& 8 seq. & 7\,h\,15\,m & 23.0\,GB & 24.5\% & 4.49 {\notsotiny $\pm$ 0.05} & 8.53 {\notsotiny $\pm$ 0.10} & 5.41 {\notsotiny $\pm$ 0.01} & 0.81 {\notsotiny $\pm$ \textbf{0.04}} & 2.86 {\notsotiny $\pm$ \textbf{0.34}} & 3.40 {\notsotiny $\pm$ 0.03} \\
\hhline{~==========}
& 2\,s & 38\,h\,19\,m & 0.3\,GB & 0.1\% & 4.32 {\notsotiny $\pm$ \textbf{0.24}} & 8.87 {\notsotiny $\pm$ \textbf{0.27}} & 5.63 {\notsotiny $\pm$ 0.03} & \textbf{1.11} {\notsotiny $\pm$ \textbf{0.04}} & \textbf{4.55} {\notsotiny $\pm$ 0.20} & \textbf{4.14} {\notsotiny $\pm$ 0.03} \\
& 4\,s & 19\,h\,39\,m & 0.7\,GB & 0.2\% & 4.54 {\notsotiny $\pm$ 0.07} & 9.15 {\notsotiny $\pm$ 0.07} & 5.74 {\notsotiny $\pm$ \textbf{0.04}} & 1.07 {\notsotiny $\pm$ 0.03} & 4.16 {\notsotiny $\pm$ 0.10} & 3.93 {\notsotiny $\pm$ 0.06} \\
& 8\,s & 12\,h\,20\,m & 1.3\,GB & 0.3\% & 4.61 {\notsotiny $\pm$ 0.10} & \textbf{9.45} {\notsotiny $\pm$ 0.12} & \textbf{5.74} {\notsotiny $\pm$ 0.02} & 1.05 {\notsotiny $\pm$ 0.02} & 4.43 {\notsotiny $\pm$ 0.13} & 3.84 {\notsotiny $\pm$ 0.08} \\
& 16\,s & 8\,h\,50\,m & 2.6\,GB & 0.8\% & 4.48 {\notsotiny $\pm$ 0.10} & 8.93 {\notsotiny $\pm$ 0.11} & 5.44 {\notsotiny $\pm$ 0.02} & 0.89 {\notsotiny $\pm$ 0.03} & 3.48 {\notsotiny $\pm$ \textbf{0.36}} & 3.62 {\notsotiny $\pm$ \textbf{0.11}} \\
& 32\,s & 10\,h\,10\,m & 5.2\,GB & 12.5\% & \textbf{4.64} {\notsotiny $\pm$ 0.04} & 9.10 {\notsotiny $\pm$ 0.13} & 5.51 {\notsotiny $\pm$ 0.03} & 0.86 {\notsotiny $\pm$ 0.00} & 3.07 {\notsotiny $\pm$ 0.09} & 3.40 {\notsotiny $\pm$ 0.02} \\
& 64\,s & 8\,h\,22\,m & 10.1\,GB & 20.3\% & 4.56 {\notsotiny $\pm$ 0.03} & 8.99 {\notsotiny $\pm$ 0.11} & 5.52 {\notsotiny $\pm$ 0.02} & 0.82 {\notsotiny $\pm$ 0.02} & 2.90 {\notsotiny $\pm$ 0.33} & 3.44 {\notsotiny $\pm$ 0.05} \\
& 128\,s & 7\,h\,18\,m & 20.2\,GB & 24.1\% & 4.54 {\notsotiny $\pm$ 0.06} & 8.60 {\notsotiny $\pm$ 0.07} & 5.41 {\notsotiny $\pm$ 0.00} & 0.80 {\notsotiny $\pm$ 0.03} & 2.70 {\notsotiny $\pm$ 0.14} & 3.40 {\notsotiny $\pm$ 0.02} \\
\hline \hline
\multirow{10}{*}{\rotatebox[origin=c]{90}{Sorted}}
& 1 seq. & 8\,h\,31\,m & 2.9\,GB & 0.0\% & 4.47 {\notsotiny $\pm$ \textbf{0.09}} & 8.81 {\notsotiny $\pm$ \textbf{0.26}} & 5.51 {\notsotiny $\pm$ 0.03} & \textbf{0.86} {\notsotiny $\pm$ \textbf{0.09}} & 2.87 {\notsotiny $\pm$ \textbf{0.63}} & \textbf{3.50} {\notsotiny $\pm$ \textbf{0.07}} \\
& 2 seq. & 9\,h\,21\,m & 5.5\,GB & 0.0\% & 4.53 {\notsotiny $\pm$ 0.03} & \textbf{8.94} {\notsotiny $\pm$ 0.07} & \textbf{5.51} {\notsotiny $\pm$ 0.02} & 0.80 {\notsotiny $\pm$ 0.03} & 2.76 {\notsotiny $\pm$ 0.16} & 3.39 {\notsotiny $\pm$ 0.03} \\
& 4 seq. & 6\,h\,58\,m & 11.0\,GB & 0.1\% & \textbf{4.63} {\notsotiny $\pm$ 0.07} & 8.91 {\notsotiny $\pm$ 0.06} & 5.51 {\notsotiny $\pm$ 0.03} & 0.85 {\notsotiny $\pm$ 0.05} & \textbf{3.34} {\notsotiny $\pm$ 0.06} & 3.45 {\notsotiny $\pm$ 0.06} \\
& 8 seq. & 5\,h\,56\,m & 21.9\,GB & 0.2\% & 4.45 {\notsotiny $\pm$ 0.05} & 8.62 {\notsotiny $\pm$ 0.12} & 5.38 {\notsotiny $\pm$ \textbf{0.03}} & 0.83 {\notsotiny $\pm$ 0.04} & 2.96 {\notsotiny $\pm$ 0.31} & 3.45 {\notsotiny $\pm$ 0.04} \\
\hhline{~==========}
& 2\,s & 34\,h\,36\,m & 0.4\,GB & 0.0\% & 4.53 {\notsotiny $\pm$ \textbf{0.12}} & 9.12 {\notsotiny $\pm$ 0.02} & 5.73 {\notsotiny $\pm$ 0.02} & 1.04 {\notsotiny $\pm$ 0.05} & 3.97 {\notsotiny $\pm$ 0.30} & \textbf{4.12} {\notsotiny $\pm$ 0.04} \\
& 4\,s & 18\,h\,41\,m & 0.7\,GB & 0.0\% & 4.68 {\notsotiny $\pm$ 0.11} & \textbf{9.41} {\notsotiny $\pm$ 0.21} & 5.70 {\notsotiny $\pm$ \textbf{0.07}} & \textbf{1.12} {\notsotiny $\pm$ 0.06} & \textbf{4.33} {\notsotiny $\pm$ 0.11} & 3.91 {\notsotiny $\pm$ 0.02} \\
& 8\,s & 11\,h\,46\,m & 1.3\,GB & 0.0\% & 4.62 {\notsotiny $\pm$ 0.03} & 9.35 {\notsotiny $\pm$ 0.11} & \textbf{5.77} {\notsotiny $\pm$ 0.04} & 1.00 {\notsotiny $\pm$ 0.06} & 3.89 {\notsotiny $\pm$ 0.29} & 3.83 {\notsotiny $\pm$ 0.04} \\
& 16\,s & 8\,h\,15\,m & 2.6\,GB & 0.1\% & 4.63 {\notsotiny $\pm$ 0.08} & 9.19 {\notsotiny $\pm$ \textbf{0.21}} & 5.57 {\notsotiny $\pm$ 0.03} & 0.80 {\notsotiny $\pm$ 0.04} & 3.07 {\notsotiny $\pm$ 0.31} & 3.66 {\notsotiny $\pm$ 0.02} \\
& 32\,s & 8\,h\,54\,m & 5.2\,GB & 0.1\% & 4.48 {\notsotiny $\pm$ 0.08} & 8.97 {\notsotiny $\pm$ 0.09} & 5.52 {\notsotiny $\pm$ 0.02} & 0.86 {\notsotiny $\pm$ \textbf{0.07}} & 3.02 {\notsotiny $\pm$ \textbf{0.37}} & 3.53 {\notsotiny $\pm$ 0.09} \\
& 64\,s & 6\,h\,45\,m & 10.1\,GB & 0.2\% & \textbf{4.76} {\notsotiny $\pm$ 0.02} & 9.10 {\notsotiny $\pm$ 0.06} & 5.54 {\notsotiny $\pm$ 0.02} & 0.92 {\notsotiny $\pm$ 0.03} & 3.19 {\notsotiny $\pm$ 0.12} & 3.48 {\notsotiny $\pm$ 0.04} \\
& 128\,s & 5\,h\,48\,m & 20.1\,GB & 0.4\% & 4.72 {\notsotiny $\pm$ 0.04} & 8.83 {\notsotiny $\pm$ 0.03} & 5.51 {\notsotiny $\pm$ 0.02} & 0.84 {\notsotiny $\pm$ 0.03} & 2.86 {\notsotiny $\pm$ 0.02} & 3.39 {\notsotiny $\pm$ \textbf{0.11}} \\
\hline \hline
\multirow{10}{*}{\rotatebox[origin=c]{90}{Bucket}}
& 1 seq. & 8\,h\,39\,m & 2.9\,GB & 0.0\% & 4.34 {\notsotiny $\pm$ \textbf{0.12}} & 8.85 {\notsotiny $\pm$ 0.08} & \textbf{5.49} {\notsotiny $\pm$ 0.02} & 0.78 {\notsotiny $\pm$ 0.03} & 2.81 {\notsotiny $\pm$ 0.20} & \textbf{3.54} {\notsotiny $\pm$ 0.05} \\
& 2 seq. & 9\,h\,28\,m & 5.8\,GB & 2.2\% & \textbf{4.59} {\notsotiny $\pm$ 0.05} & \textbf{8.95} {\notsotiny $\pm$ 0.03} & 5.46 {\notsotiny $\pm$ 0.03} & \textbf{0.88} {\notsotiny $\pm$ 0.05} & \textbf{3.18} {\notsotiny $\pm$ 0.06} & 3.50 {\notsotiny $\pm$ 0.02} \\
& 4 seq. & 7\,h\,11\,m & 11.6\,GB & 4.0\% & 4.51 {\notsotiny $\pm$ 0.05} & 8.87 {\notsotiny $\pm$ \textbf{0.15}} & 5.43 {\notsotiny $\pm$ \textbf{0.03}} & 0.86 {\notsotiny $\pm$ \textbf{0.07}} & 2.97 {\notsotiny $\pm$ \textbf{0.35}} & 3.41 {\notsotiny $\pm$ 0.01} \\
& 8 seq. & 6\,h\,13\,m & 23.0\,GB & 5.2\% & 4.40 {\notsotiny $\pm$ 0.05} & 8.59 {\notsotiny $\pm$ 0.05} & 5.43 {\notsotiny $\pm$ 0.00} & 0.74 {\notsotiny $\pm$ 0.03} & 2.66 {\notsotiny $\pm$ 0.13} & 3.42 {\notsotiny $\pm$ \textbf{0.09}} \\
\hhline{~==========}
& 2\,s & 36\,h\,12\,m & 0.3\,GB & 0.2\% & 4.65 {\notsotiny $\pm$ \textbf{0.12}} & 9.46 {\notsotiny $\pm$ \textbf{0.15}} & 5.73 {\notsotiny $\pm$ 0.01} & \textbf{1.16} {\notsotiny $\pm$ 0.05} & \textbf{4.84} {\notsotiny $\pm$ 0.10} & \textbf{4.07} {\notsotiny $\pm$ 0.06} \\
& 4\,s & 20\,h\,08\,m & 0.7\,GB & 0.3\% & \textbf{4.77} {\notsotiny $\pm$ 0.07} & \textbf{9.58} {\notsotiny $\pm$ 0.04} & \textbf{5.81} {\notsotiny $\pm$ \textbf{0.02}} & 1.05 {\notsotiny $\pm$ 0.05} & 4.12 {\notsotiny $\pm$ 0.26} & 3.97 {\notsotiny $\pm$ 0.07} \\
& 8\,s & 12\,h\,23\,m & 1.3\,GB & 0.4\% & 4.70 {\notsotiny $\pm$ 0.04} & 9.46 {\notsotiny $\pm$ 0.13} & 5.78 {\notsotiny $\pm$ 0.02} & 1.00 {\notsotiny $\pm$ \textbf{0.06}} & 4.16 {\notsotiny $\pm$ \textbf{0.43}} & 3.80 {\notsotiny $\pm$ 0.06} \\
& 16\,s & 8\,h\,26\,m & 2.6\,GB & 0.6\% & 4.60 {\notsotiny $\pm$ 0.02} & 9.12 {\notsotiny $\pm$ 0.09} & 5.57 {\notsotiny $\pm$ 0.01} & 0.92 {\notsotiny $\pm$ 0.06} & 3.65 {\notsotiny $\pm$ 0.09} & 3.69 {\notsotiny $\pm$ \textbf{0.11}} \\
& 32\,s & 8\,h\,38\,m & 4.6\,GB & 1.7\% & 4.74 {\notsotiny $\pm$ 0.10} & 9.25 {\notsotiny $\pm$ 0.06} & 5.58 {\notsotiny $\pm$ 0.01} & 0.88 {\notsotiny $\pm$ 0.04} & 3.33 {\notsotiny $\pm$ 0.20} & 3.60 {\notsotiny $\pm$ 0.01} \\
& 64\,s & 7\,h\,16\,m & 10.1\,GB & 4.0\% & 4.67 {\notsotiny $\pm$ 0.07} & 9.03 {\notsotiny $\pm$ 0.13} & 5.53 {\notsotiny $\pm$ 0.02} & 0.90 {\notsotiny $\pm$ 0.05} & 3.38 {\notsotiny $\pm$ 0.36} & 3.52 {\notsotiny $\pm$ 0.03} \\
& 128\,s & 6\,h\,10\,m & 20.1\,GB & 5.2\% & 4.56 {\notsotiny $\pm$ 0.05} & 9.02 {\notsotiny $\pm$ 0.01} & 5.51 {\notsotiny $\pm$ 0.01} & 0.78 {\notsotiny $\pm$ 0.02} & 3.05 {\notsotiny $\pm$ 0.17} & 3.19 {\notsotiny $\pm$ 0.04} \\
\hline \hline
\end{tabular}
\caption{Training statistics and speech enhancement scores in matched and mismatched conditions for different batching strategies.}
\label{tab:batching}
\end{table*}

For each strategy, using a fixed batch size of 8 sequences or a dynamic batch size of 128\,s results in similar training times. This is in line with the sequence length distribution in Fig.\,\ref{fig:dist}, which has its peak around 16\,s, meaning 8 random sequences can be expected to correspond to 128\,s in total. While the training times are similar, the allocated GPU memory is greater when using fixed 8 sequences-long batches. We argue this is due to the inconsistency of the total amount of data in batches of fixed length; when using a fixed batch size, batches of short mixtures consist of significantly fewer audio samples compared to batches of long mixtures. This highlights the benefit of using a dynamic batch size over a fixed batch size when using highly variable size inputs.

In Fig.~\ref{fig:results}, the speech enhancement scores in matched and mismatched conditions are shown as a function of the dynamic batch size. It can be seen that the performance in terms of $\destoi$ and $\dsnr$ decreases with larger batch sizes for all three batching strategies. This trend is significantly stronger in mismatched conditions. No particular trend can be observed for $\dpesq$. This is in line with \cite{keskar2016large} and the common assumption that smaller batch sizes can result in models with better generalization. Interestingly, sorted batching provides similar performance compared to random batching. This suggests that the training does not benefit significantly from randomizing the data in each batch, and shuffling the batches at the start of each epoch is sufficient. Sorted batching thus appears as a solution for minimum padding and improved training times at seemingly no compromise on speech enhancement performance compared to random batching. This is in line with recent studies questioning the need for i.i.d. sampling of the training data \cite{meng2017convergence, nguyen2022why}, although they were in the context of distributed optimization.

\section{Conclusion}
\label{sec:conclusion}

We conducted an empirical study of three batching strategies for the training of an end-to-end speech enhancement system with variable size noisy and reverberant mixtures. Specifically, we investigated how batching influences training times, GPU memory allocation and speech enhancement performance in both matched and mismatched conditions. We found that a smaller batch size provides better performance for all batching strategies, especially in mismatched conditions. Moreover, using a dynamic batch size for a more consistent total amount of data in each batch showed to reduce the allocated GPU memory at equal training times. Finally, using sorted or bucket batching allowed for substantially lower \gls{zpr} and thereby improved training times without compromising performance, which indicates that shuffling the batches before each epoch is sufficient to obtain appropriate gradient estimates. While the present study focused on speech enhancement, we believe it can serve as a guideline for future researchers who need to develop data loading pipelines for variable size inputs of different nature as well.

\bibliographystyle{IEEEtran}
\bibliography{IEEEabrv, abbrv, refs}

\end{document}